\documentclass[conference,a4paper]{APSIPA2026}
\usepackage{amsmath}
\usepackage[dvips,dvipdfm]{graphicx}

\usepackage{multirow}
\usepackage{threeparttable}

\usepackage{subcaption}
\usepackage{siunitx}
\usepackage{booktabs}
\usepackage{bm}
\usepackage{amsmath,amssymb}
\usepackage{url}

\usepackage{geometry}
\usepackage{fancyhdr}

\fancypagestyle{firststyle}{
  \fancyhf{}
  \fancyhead[C]{}
}

\begin{document}

\title{Audio-Image Cross-Modal Retrieval\\with Onomatopoeic Images}

\author{
\authorblockN{
Keisuke Imoto\authorrefmark{2}, Yamato Kojima\authorrefmark{3} and Takao Tsuchiya\authorrefmark{3}\\[3pt]
}

\authorblockA{
\authorrefmark{2}Kyoto University, Japan, \ \ \ \authorrefmark{3}Doshisha University, Japan}
}

\maketitle
\thispagestyle{firststyle}
\pagestyle{empty}

\begin{abstract}
Finding sound effects or environmental sounds that match a creator's intended impression remains a largely manual process in multimedia production.
This is especially relevant for comics and other visual media, where visually stylized onomatopoeic expressions convey auditory impressions through letter shapes, strokes, layouts, and decorative patterns.
However, cross-modal retrieval between onomatopoeic images and general sounds has been largely unexplored.
This paper thus introduces a bidirectional retrieval framework between onomatopoeic images and the corresponding sound clips.
Instead of directly comparing embeddings extracted from pretrained image and audio encoder, we train modality-specific projection heads that re-align the embeddings for visual onomatopoeia and corresponding sounds.
We then construct the Multimodal Image-Audio Onomatopoeia dataset (MIAO), which contains paired onomatopoeic images and sound clips across 50 sound event classes.
Experimental results show that the proposed method substantially outperforms a zero-shot baseline using pretrained CLIP and CLAP embeddings.
These results demonstrate that adapting pretrained representations enables effective retrieval in both directions: from onomatopoeic images to sounds and from sounds to onomatopoeic images.
%
%
\end{abstract}

\section{Introduction}
Sound effect design plays an essential role in multimedia content production, including animation, games, and film.
In practice, creators often search for and select appropriate sounds based on their intended auditory impression.
However, this process is largely manual and relies heavily on individual experience, making it difficult to efficiently explore a wide range of candidate sounds while maintaining consistency in expression.

A natural way to support this process is to use visual cues that express the intended sound impression.
In comics and illustrated media, such cues often appear as visually stylized onomatopoeic expressions, whose letter shapes, strokes, layouts, and decorative patterns can reflect sound-related information.
This makes onomatopoeic images a promising cue for the corresponding sound retrieval.
Conversely, sound-to-image retrieval allows us to examine which visual onomatopoeic expressions are associated with a given sound.
Nevertheless, cross-modal retrieval between onomatopoeic images and general sounds, including sound effects and environmental sounds, has not been sufficiently explored.

To address this gap, we propose a cross-modal representation learning method for onomatopoeic image--audio retrieval.
The proposed method builds on pretrained image and audio encoders and trains lightweight modality-specific projection heads that map onomatopoeic images and audios into a shared embedding space.
This design allows us to exploit pretrained visual and audio representations while adapting them to the correspondence between visual onomatopoeia and corresponding sounds.
Experimental results show that the proposed method substantially outperforms a zero-shot baseline that directly compares pretrained Contrastive Language–Image Pre-training (CLIP) \cite{Radford_ICML2021_01} and Contrastive Language-Audio Pre-training (CLAP) \cite{Wu_ICASSP2023_01} embeddings, demonstrating the importance of adapting pretrained representations for onomatopoeic image--audio retrieval.

The remainder of this paper is organized as follows.
Section~3 describes the proposed cross-modal representation learning method.
Section~4 presents the dataset, experimental setup, evaluation metrics, and retrieval results.
Finally, Section~5 concludes the paper and discusses future work.
%
%
\section{Related Work}
Large-scale cross-modal models provide a useful basis for retrieval across modalities.
For example, CLIP trains a shared embedding space for images and text from large-scale image--caption pairs~\cite{Radford_ICML2021_01}, while CLAP learns a shared space for audio and text from audio--caption pairs~\cite{Wu_ICASSP2023_01}.
However, these spaces are learned independently, and thus, CLIP image embeddings and CLAP audio embeddings are not guaranteed to be directly comparable.
To address this limitation, several studies have trained explicit relationship between audio and visual representations, including AudioCLIP~\cite{Guzhov_ICASSP2022_01}, Wav2CLIP~\cite{Wu_ICASSP2022_01}, and ImageBind~\cite{Girdhar_CVPR2023_01}.

Onomatopoeia has been studied as a compact expression for describing fine-grained characteristics of sounds.
Ikawa and Kashino proposed sound event search using onomatopoeic words by mapping sounds and onomatopoeic text into a common latent space~\cite{Ikawa_DCASE2018_01}.
Onomatopoeic words have also been used for environmental sound synthesis, as in Onoma-to-wave~\cite{Okamoto_ATSIP2021_01}.
These studies show that onomatopoeia is useful for representing fine-grained sound characteristics, but they treat only onomatopoeia text sequences.

\begin{figure*}[t]
  \centering
  \includegraphics[width=0.85\linewidth]{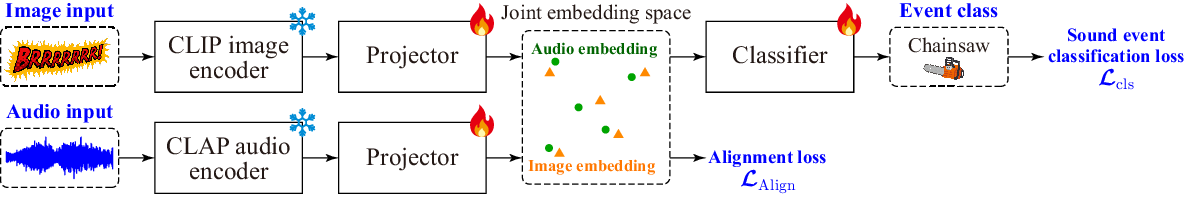}
  \vspace{2pt}
  \caption{Summary of proposed onomatopoeic image--audio representation learning.}
  \label{fig:proposed_learning}
\end{figure*}

Visual onomatopoeia has also been explored as a bridge from sounds to visual effects.
Wang et al.~\cite{Wang_TMM2017_01} studied animated sound words for visualizing nonverbal sounds in videos.
Their work controls how font-based onomatopoeic words are presented as visual effects.
However, it does not fully address the sound-expressive information inherent in onomatopoeic images, such as freely drawn letter shapes, strokes, layouts, and decorative patterns.
Visual onoma-to-wave~\cite{Ohnaka_ICASSP2023_01} synthesizes environmental sounds from visually represented onomatopoeias and sound-source images.

In contrast, this paper studies bidirectional retrieval between onomatopoeic images and general/environmental sounds.
We train an aligned embedding space that connects visual onomatopoeic expressions with corresponding sound clips, enabling retrieval from images to sounds and from sounds to images.

\color{black}
%
%
%
\section{Proposed Method}
Although large-scale pretrained models such as CLIP and CLAP provide strong general representations, visual onomatopoeic images are rarely included in their pretraining data.
Moreover, the onomatopoeic images differ from natural images in how they convey information, expressing auditory impressions through letter shapes, strokes, spatial layouts, and decorative patterns.
Therefore, directly comparing pretrained image and audio embeddings is unlikely to capture the correspondence between visual onomatopoeia and general sounds.
To address this mismatch, we introduce modality-specific projection heads on top of pretrained image and audio encoders.
\begin{figure*}[t]
  \vspace*{5pt}
  \centering
  \includegraphics[width=0.70\linewidth]{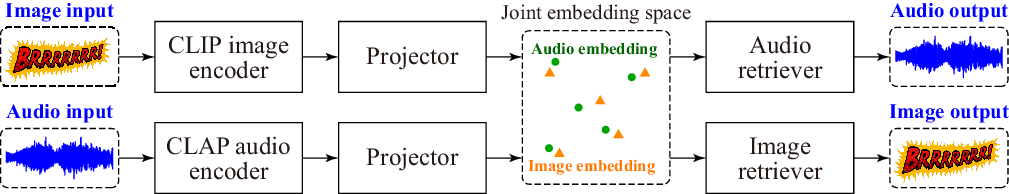}
  \vspace{2pt}
  \caption{Summary of proposed onomatopoeic image--audio retriever.}
  \label{fig:proposed_retriever}
\end{figure*}
%
%
\subsection{Cross-modal representation learning for visual onomatopoeia and sound}
To align embeddings of onomatopoeic images and corresponding sounds, the proposed method introduces projection heads on top of embeddings extracted from the image and audio encoders, as shown in Fig.~\ref{fig:proposed_learning}.
The pretrained image and audio encoders already provide a general shared embedding space, however this space is not specialized for the visual cues of onomatopoeic images.
The projection heads thus re-align the pretrained embeddings toward the correspondence between visual onomatopoeia and corresponding sounds.

Let $\bm{x}_{\mathrm{img}}$ and $\bm{x}_{\mathrm{aud}}$ denote an onomatopoeic image and a corresponding sound, respectively.
We first obtain modality-specific embeddings using the image and audio encoders:
\begin{align}
\bm{z}_{\mathrm{img}} &= \mathcal{F}_{\mathrm{img}}(\bm{x}_{\mathrm{img}}), \\
\bm{z}_{\mathrm{aud}} &= \mathcal{F}_{\mathrm{aud}}(\bm{x}_{\mathrm{aud}}),
\end{align}
where $\mathcal{F}_{\mathrm{img}}(\cdot)$ and $\mathcal{F}_{\mathrm{aud}}(\cdot)$ denote the image and audio encoders, respectively.
The extracted embeddings are
$\bm{z}_{\mathrm{img}}, \bm{z}_{\mathrm{aud}} \in \mathbb{R}^{D}$.
In this work, we instantiate these encoders with the CLIP image encoder and the CLAP audio encoder, but the proposed approach can be used with other image and audio encoders.

The modality-specific embeddings $\bm{z}_{\mathrm{img}}$ and $\bm{z}_{\mathrm{aud}}$ are then projected into a joint embedding space:
\begin{align}
\tilde{\bm{z}}_{\mathrm{img}}
&= \mathcal{G}_{\mathrm{img}}(\bm{z}_{\mathrm{img}}), \\
\tilde{\bm{z}}_{\mathrm{aud}}
&= \mathcal{G}_{\mathrm{aud}}(\bm{z}_{\mathrm{aud}}),
\end{align}
where $\mathcal{G}_{\mathrm{img}}(\cdot)$ and $\mathcal{G}_{\mathrm{aud}}(\cdot)$ are the image and audio projection heads.
The projected embeddings
$\tilde{\bm{z}}_{\mathrm{img}}, \tilde{\bm{z}}_{\mathrm{aud}} \in \mathbb{R}^{\tilde{D}}$
form an aligned joint embedding space for onomatopoeic images and the corresponding sounds.
During training, the projection heads are trained so that paired or class-consistent onomatopoeic images and sound clips become close in this embedding space, while preserving sound event discriminability.

In our implementation, the pretrained CLIP and CLAP encoders are kept frozen, and only the projection heads and the classifier are updated during the model training stage.
Each projection head is implemented as two fully-connected layers.

To preserve sound event information in the joint embedding space, we apply a common classifier to the projected embeddings during training:
\begin{align}
\bm{s}_{\mathrm{img}} &= \mathcal{H}(\tilde{\bm{z}}_{\mathrm{img}}), \\
\bm{s}_{\mathrm{aud}} &= \mathcal{H}(\tilde{\bm{z}}_{\mathrm{aud}}),
\end{align}
where $\mathcal{H}(\cdot)$ denotes the classifier, and $\bm{s}_{\mathrm{img}}, \bm{s}_{\mathrm{aud}} \in \mathbb{R}^{C}$ are class score vectors over $C$ sound event classes.
The classifier is used only during training stage and is removed during retrieval stage.
%
%
\subsection{Training objective}
The projection heads are trained so that the joint embedding space satisfies two requirements.
First, an onomatopoeic image and a corresponding sound embeddings should be placed close to each other.
Second, the trained embeddings should retain class discriminative information.
To this end, we train the proposed cross-modal model with and alignment loss and a sound event classification loss.

For the alignment loss, we minimize the squared Euclidean distance between their projected embeddings:
\begin{align}
\mathcal{L}_{\mathrm{align}} =
\left\lVert
\tilde{\bm{z}}_{\mathrm{img}} - \tilde{\bm{z}}_{\mathrm{aud}}
\right\rVert_2^2 .
\label{eq:loss_pair}
\end{align}
This loss directly encourages the projected image and audio embeddings from the same pair to be close in the joint embedding space.

For the sound event classification loss, we apply a cross-entropy loss to the class scores produced from both modalities.
Given the ground-truth sound event label $y$, the classification loss is defined as
\begin{align}
\mathcal{L}_{\mathrm{cls}} = \mathrm{CE}\!\left(\bm{s}_{\mathrm{img}}, y\right) + \mathrm{CE}\!\left(\bm{s}_{\mathrm{aud}}, y\right),
\label{eq:loss_cls}
\end{align}
where $\mathrm{CE}(\cdot)$ denotes the cross-entropy loss.
This loss encourages projected embeddings from both onomatopoeic images and sound clips to remain discriminative with respect to sound event classes.

The total training objective is
\begin{align}
\mathcal{L} = \mathcal{L}_{\mathrm{align}} + \mathcal{L}_{\mathrm{cls}}.
\label{eq:loss_total}
\end{align}
During training, the pretrained image and audio encoders are frozen, and the parameters of the projection heads and classifier are updated by minimizing $\mathcal{L}$.
%
%
\subsection{Cross-modal retrieval}
After model training, retrieval is performed using the aligned joint embedding space obtained by the projection heads as shown in Fig.~\ref{fig:proposed_retriever}.
In the retrieval stage, the classifier is discarded, and only the pretrained encoders and the projection heads are used to obtain the embeddings.

For image-to-audio retrieval (I2A), an input onomatopoeic image $\bm{x}_{\mathrm{img}}$ is encoded and projected as
\begin{align}
\tilde{\bm{z}}_{\mathrm{img}} = \mathcal{G}_{\mathrm{img}}
\left(
\mathcal{F}_{\mathrm{img}}(\bm{x}_{\mathrm{img}})
\right).
\end{align}
Each audio candidate $\bm{x}_{\mathrm{aud},j}$ is also encoded and projected as
\begin{align}
\tilde{\bm{z}}_{\mathrm{aud},j}
=
\mathcal{G}_{\mathrm{aud}}
\left(
\mathcal{F}_{\mathrm{aud}}(\bm{x}_{\mathrm{aud},j})
\right).
\end{align}
The retrieval score is then calculated by cosine similarity:
\begin{align}
\mathrm{sim}( \tilde{\bm{z}}_{\mathrm{img}}, \tilde{\bm{z}}_{\mathrm{aud},j} ) =
\frac{(\tilde{\bm{z}}_{\mathrm{img}})^\top \tilde{\bm{z}}_{\mathrm{aud},j} }{
\left\lVert \tilde{\bm{z}}_{\mathrm{img}} \right\rVert_2
\left\lVert \tilde{\bm{z}}_{\mathrm{aud},j} \right\rVert_2
}.
\label{eq:retrieval_score}
\end{align}
Audio candidates are ranked in descending order of this score.

Audio-to-image retrieval (A2I) is performed in the same joint embedding space by reversing the query and candidate modalities.
An sound clip is used as the query, and onomatopoeic images are ranked according to their cosine similarity to the projected audio embedding.
Thus, the same trained space supports bidirectional retrieval between onomatopoeic images and sound clips.
\begin{figure}[t]
  \centering
  \begin{subfigure}[t]{0.47\linewidth}
    \centering
    \vspace{-70pt}
    \includegraphics[width=\linewidth]{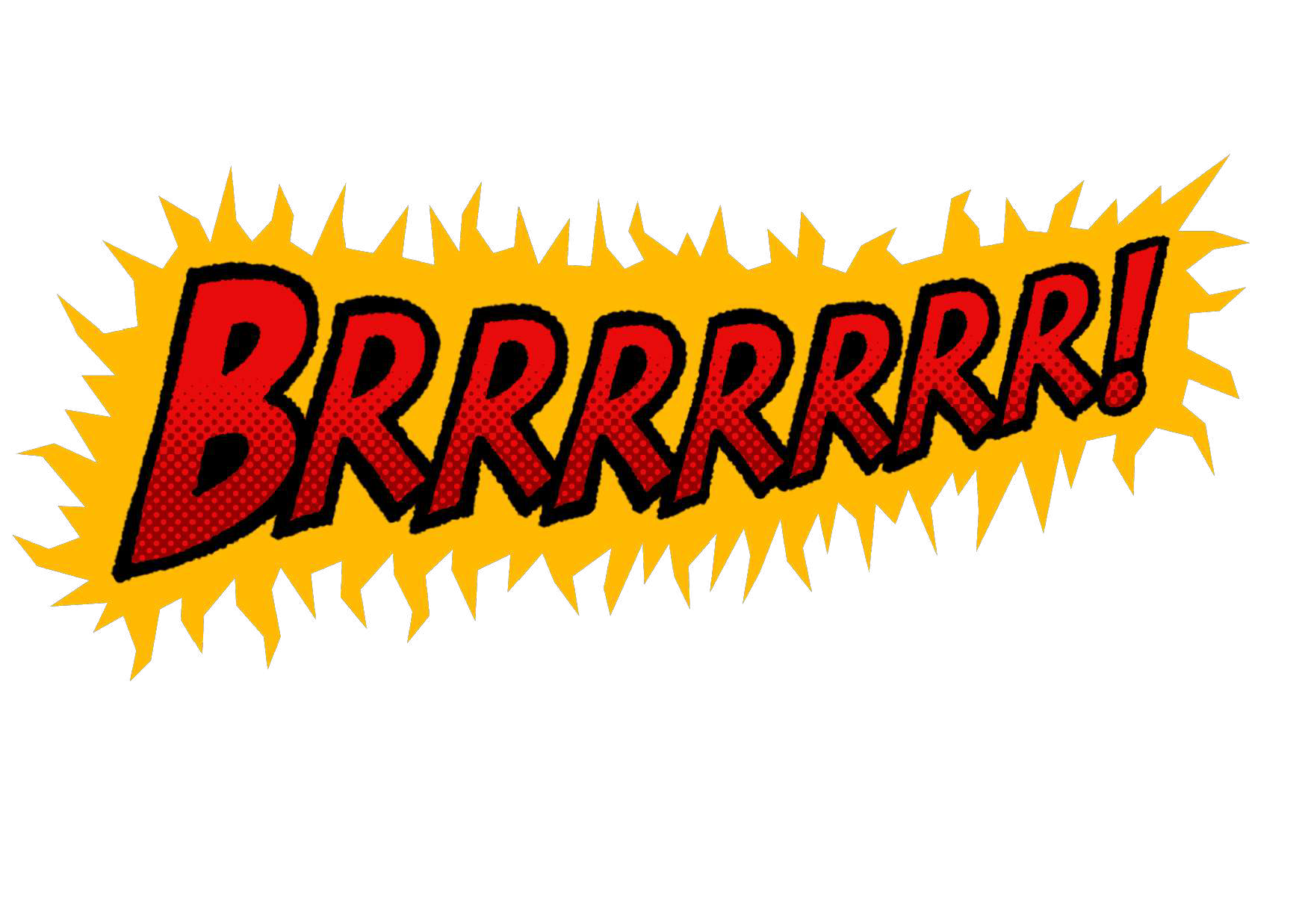}
    \vspace{-29pt}
    \caption{Onomatopoeic image of ``chainsaw''}
  \end{subfigure}\hfill
  \hspace{10pt}
  \begin{subfigure}[t]{0.47\linewidth}
    \centering
    \includegraphics[width=\linewidth]{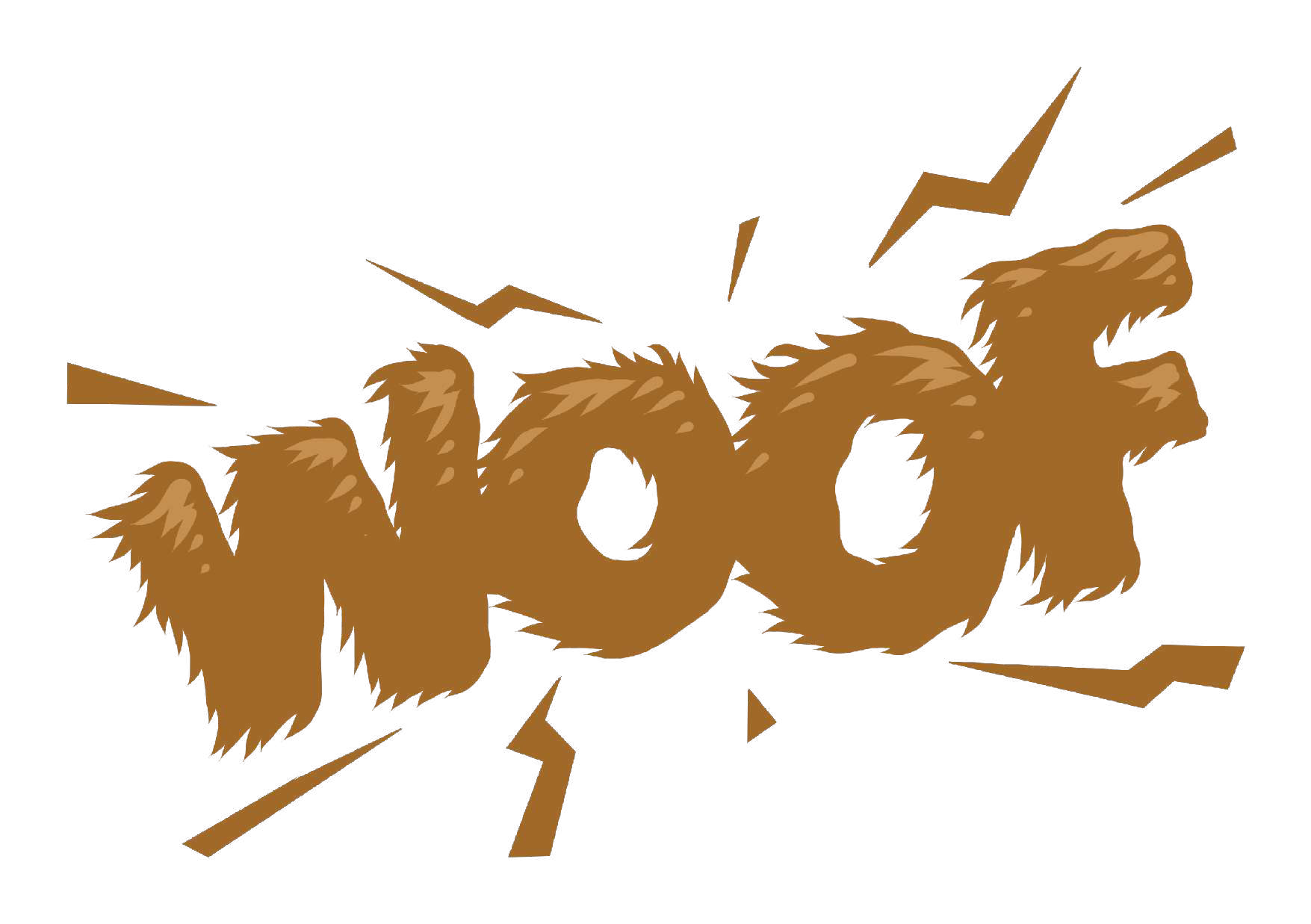}
    \vspace{-15pt}
    \caption{Onomatopoeic image of ``dog barking''}
    \end{subfigure}
  \caption{Examples of onomatopoeic images in MIAO dataset}
  \label{fig:ono_image_samples}
\end{figure}
%
%
\section{Experimental Evaluation}
To evaluate whether the proposed method can correctly associate visual onomatopoeic expressions with environmental sounds, we conducted retrieval experiments between onomatopoeic images and sound clips.
As a comparison method, we used a zero-shot baseline that directly compares embeddings extracted from pretrained CLIP and CLAP encoders.
Retrieval was evaluated in both directions: image-to-audio retrieval (I2A), where an onomatopoeic image is used as a query, and audio-to-image retrieval (A2I), where an sound clip is used as a query.
%
%
\subsection{Dataset}
We use the Multimodal Image-Audio Onomatopoeia dataset (MIAO)\footnote{\url{https://huggingface.co/datasets/KeisukeImoto/MIAO}}, a newly constructed dataset that pairs visual onomatopoeic expressions with environmental sounds.
MIAO consists of 850 image--audio pairs across 50 sound event classes.
Each pair contains a sound clip and an onomatopoeic image drawn after listening to the sound clip.
The images were created by 17 illustrators, providing a variety of visual styles even for the same sound class.
The sound clips were selected from the Creative Commons Zero (CC0) subset of FSD50K~\cite{Fonseca_TASLP2022_01}.
Figure~\ref{fig:ono_image_samples} shows examples of onomatopoeic images in MIAO.
%
%
\subsection{Experimental setup}
To evaluate generalization to unseen visual styles, MIAO was split by illustrator.
The training set consisted of 650 image--audio pairs from 13 illustrators, and the validation/test set consisted of 100 pairs from two illustrators, respectively.

For the zero-shot baseline, each onomatopoeic image was encoded using the CLIP image encoder, and each sound clip was encoded using the CLAP audio encoder.
In our implementation, the CLIP image encoder uses a ViT-B/32 backbone \cite{Radford_ICML2021_01}, and the CLAP audio encoder uses an HTS-AT backbone \cite{Chen_ICASSP2022_01}.
The resulting 512-dimensional embeddings were $\ell_2$-normalized and ranked by cosine similarity.
This baseline evaluated whether pretrained image and audio representations could be directly compared for onomatopoeic image--sound retrieval without task-specific adaptation.

For the proposed method, the CLIP image encoder and CLAP audio encoder were kept frozen throughout training.
We trained only the image projector, audio projector, and classifier.
Each modality-specific projector was implemented as a two layer MLP projection head with dimensions $512 \rightarrow 512 \rightarrow 256$.
The classifier mapped the 256-dimensional shared embedding to $C=50$ sound event classes.

The model was trained using AdamW~\cite{Loshchilov_ICLR2019_01} with a learning rate of $1.0 \times 10^{-3}$ and a weight decay of $1.0 \times 10^{-4}$.
The dropout rate was set to 0.1 and the batch size to 64.
All experiments were conducted with ten random seeds, and the results are reported as the mean and unbiased standard deviation.
\begin{table}[t]
\scriptsize
\centering
\caption{Retrieval performance of zero-shot baseline and proposed method}
\label{tab:main_results}
    \begin{tabular}{llccccc}
    \toprule
    \multicolumn{2}{c}{\textbf{Method}}\!&\!\textbf{mAP (\%)}\!&\!\textbf{R@1 (\%)}\!&\!\textbf{R@5 (\%)}\!&\!\textbf{MRR} \\
    \midrule
    \multirow{2}{*}{\textbf{Baseline}} \!\!\!\!\!\!\!
    & \textbf{I2A}\!&\!6.77 $\pm$ 0.00\!&\!2.00 $\pm$ 0.00\!&\!\phantom{0}9.00 $\pm$ 0.00\!&\!0.076 $\pm$ 0.00 \\
    & \textbf{A2I}\!&\!7.82 $\pm$ 0.00\!&\!6.00 $\pm$ 0.00\!&\!10.00 $\pm$ 0.00\!&\!0.116 $\pm$ 0.00 \\
    \midrule
    \multirow{2}{*}{\textbf{Proposed}} \!\!\!\!\!\!\!
    & \textbf{I2A}\!&\!61.45 $\pm$ 1.71\!&\!53.60 $\pm$ 2.41\!&\!68.90 $\pm$ 2.69\!&\!0.60 $\pm$ 0.02 \\
    & \textbf{A2I}\!&\!61.08 $\pm$ 1.84\!&\!64.60 $\pm$ 3.37\!&\!88.20 $\pm$ 2.66\!&\!0.75 $\pm$ 0.03 \\
    \bottomrule
    \end{tabular}
\end{table}
%
%
\subsection{Evaluation metrics}
We use mean average precision (mAP), Recall@$k$ (R@$k$), and mean reciprocal rank (MRR) as the evaluation metrics.
For retrieval, one modality is used as a query and the corresponding samples in the other modality are ranked according to their similarity to the query.
A retrieved item is considered correct if it belongs to the same event class as the query.

In MIAO, each event class is represented by multiple sound clips, and each sound clip is paired with onomatopoeic images drawn by multiple illustrators.
Therefore, retrieval is also evaluated at the class level; that is, the correct item is not limited to the original paired sample, and other samples from the same event class are also treated as correct.
mAP is used to evaluate whether all correct items for a query are consistently placed near the top of the ranked list, which is important in our class-level evaluation with multiple correct items.

R@1 and R@5 are used to evaluate whether the retrieval system can return at least one correct item within a small number of candidates.

MRR is used to evaluate how quickly the system reaches the first correct item by penalizing cases where the first correct item appears only after several incorrect ones.
\newcommand{\hs}{\hspace{1.2mm}}
\begin{table}[t]
\small
\centering
\caption{Five lowest-performing classes in image-to-audio retrieval, measured by class-wise AP}
\label{tab:bottom5_ap_confusion_i2a}
    \begin{tabular}{clcl}
        \toprule
        Rank & Sound event & AP [\%] & Most confused class\\
        \midrule
        1 & Camera    & \hs 6.10 $\pm$ 4.30 & Keyboard typing \\
        2 & Boiling   & \hs 7.60 $\pm$ 3.40 & Thunder \\
        3 & Sea waves & \hs 9.10 $\pm$ 2.60 & Wind \\
        4 & Train     & 14.10 $\pm$ 9.20 & Cymbal \\
        5 & Drill     & 15.40 $\pm$ 8.50 & Frying food \\
        \bottomrule
    \end{tabular}
\end{table}
\begin{table}[t]
\small
\centering
\caption{Five lowest-performing classes in audio-to-image retrieval, measured by class-wise AP}
\label{tab:bottom5_ap_confusion_a2i}
    \begin{tabular}{clcl}
        \toprule
        Rank & Sound event & AP [\%] & Most confused class\\
        \midrule
        1 & Camera    & \hs 8.10 $\pm$ 6.40  & Scissors \\
        2 & Boiling   & \hs 8.20 $\pm$ 4.10  & Pouring liquid \\
        3 & Train     & 12.30 $\pm$ 8.00 & Aircraft \\
        4 & Sea waves & 15.50 $\pm$ 6.60 & Glass breaking \\
        5 & Drill     & 17.30 $\pm$ 8.40 & Chainsaw \\
        \bottomrule
    \end{tabular}
\end{table}
\begin{table}[t]
\centering
\caption{Average cosine distance from class centroid for lowest-performing 5 classes}
\label{tab:dispersion_bottom}
\small
  \begin{tabular}{lcc}
  \toprule
  Sound event & Audio dispersion & Image dispersion \\
  \midrule
  Camera & 0.0008 $\pm$ 0.0009 & 0.2116 $\pm$ 0.0665 \\
  Boiling & 0.0007 $\pm$ 0.0003 & 0.3009 $\pm$ 0.0309 \\
  Sea waves & 0.0007 $\pm$ 0.0003 & 0.3017 $\pm$ 0.0281 \\
  Train & 0.0270 $\pm$ 0.0100 & 0.1653 $\pm$ 0.0442 \\
  Drill & 0.0122 $\pm$ 0.0052 & 0.2836 $\pm$ 0.0466 \\
  \bottomrule
  \end{tabular}
\end{table}
%
%
\subsection{Experimental results}
Table~\ref{tab:main_results} shows the retrieval performance of the zero-shot baseline and the proposed method.
Note that the zero-shot baseline has no trainable components, so its results are deterministic across runs.
The proposed method substantially outperforms the baseline in both image-to-audio retrieval and audio-to-image retrieval.
In image-to-audio retrieval, the proposed method improves mAP from 6.77\% to 61.45\%, and R@1 from 2.00\% to 53.60\%.
In audio-to-image retrieval, mAP improves from 7.82\% to 61.08\%, and R@1 improves from 6.00\% to 64.60\%.
These results indicate that directly comparing pretrained CLIP and CLAP embeddings is insufficient for onomatopoeic image--audio retrieval, whereas the proposed projection heads effectively re-align the embeddings for visual onomatopoeia and the corresponding sound events.

A closer look at the top-ranked results shows that the improvement is not limited to the first retrieved item.
When multiple top-ranked items are considered, the proposed method also achieves large gains in R@5.
In image-to-audio retrieval, R@5 reach 68.90\%, and  in audio-to-image retrieval, it further increase to 88.20\%.
The MRR values also improve substantially over the baseline, indicating that the first correct item appears much earlier in the ranked list.
These results suggest that the proposed method improves not only the top-1 retrieval accuracy but also the consistency of the upper-ranked retrieval results.

The results also reveal a difference in performance between the two retrieval directions.
Although image-to-audio and audio-to-image retrieval achieve similar mAP values, audio-to-image retrieval obtains higher R@1, R@5, and MRR.
This indicates that, in the proposed embedding space, an audio query more easily retrieves at least one correct onomatopoeic image among the top-ranked results than an onomatopoeic image query retrieves correct sounds.
One possible reason is that onomatopoeic images exhibit larger visual variation than sound clips within the same sound event class, making image queries more sensitive to illustrator-dependent expressions.

Tables~\ref{tab:bottom5_ap_confusion_i2a} and~\ref{tab:bottom5_ap_confusion_a2i} show the five classes with the lowest AP values for image-to-audio and audio-to-image retrieval, respectively.
The same classes, including ``Camera,'' ``Boiling,'' ``Sea waves,'' ``Train,'' and ``Drill,'' appear among the lowest-performing classes in both directions.
This suggests that the remaining errors are concentrated in classes whose visual or audio cues are difficult to distinguish from those of other classes.
For example, ``Boiling'' is often confused with ``Pouring liquid,'' which may reflect similar continuous and bubbling sounds.
Similarly, ``Drill'' is confused with ``Chainsaw,'' likely because both classes involve mechanical and continuous sounds that can lead to visually similar onomatopoeic expressions.

To further examine the source of these errors, Table~\ref{tab:dispersion_bottom} reports the average cosine distance from the class centroid for the lowest-performing classes.
These classes show much larger dispersion in image embeddings than in audio embeddings.
For example, ``Boiling'' and ``Sea waves'' have audio dispersion values on the order of $10^{-3}$, whereas their image dispersion values are around 0.30.
This indicates that the sound clips in these classes are mapped consistently, while the corresponding onomatopoeic images vary substantially across illustrators.
Such visual variability makes it difficult to form compact class clusters in the joint embedding space and can lead to retrieval errors, especially when an onomatopoeic image is used as the query.

Overall, the results demonstrate that the proposed method effectively adapts pretrained embeddings to onomatopoeic image--audio retrieval.
At the same time, the class-wise analysis shows that the remaining errors are strongly related to the diversity of visual onomatopoeic expressions, suggesting that robust modeling of illustrator-dependent variation is an important direction for future work.
%
%
\section{Conclusion}
This paper presented a cross-modal retrieval framework between onomatopoeic images and general sounds.
To adapt pretrained image and audio embeddings to visual onomatopoeia, the proposed method introduced modality-specific projection heads on top of pretrained image and audio encoders.
The projection heads re-align the pretrained embeddings toward the correspondence between onomatopoeic images and sound events, enabling bidirectional retrieval between images and sound clips.
We also constructed MIAO, a dataset of paired visual onomatopoeic images and general sounds, and evaluated the proposed method on image-to-audio and audio-to-image retrieval.
Experimental results showed that the proposed method substantially outperformed a zero-shot baseline that directly compares pretrained CLIP and CLAP embeddings.
These results demonstrate that adapting pretrained embeddings is essential for retrieving sounds from onomatopoeic images and retrieving onomatopoeic images from sounds.
Future work needs to be conducted to explore more robust representation learning for diverse visual onomatopoeic expressions, including the use of textual onomatopoeia or sound event descriptions as additional supervision.
%
%
\section*{Acknowledgment}
This work was supported by JSPS KAKENHI Grant Numbers 22H03639 and the Hoso Bunka Foundation.

\bibliographystyle{IEEEtran}
\bibliography{APSIPA2026}

\end{document}